\documentclass[11pt,aps,tightenlines,nofootinbib,longbibliography,superscriptaddress]{revtex4-1}

\usepackage{graphicx}
\usepackage[margin=1in]{geometry}
\usepackage[usenames,dvipsnames]{color}
\usepackage{url}
\usepackage[colorlinks = true,
            linkcolor = blue,
            urlcolor  = blue,
            citecolor = blue,
            anchorcolor = blue]{hyperref}
\usepackage{setspace}

\newcommand\snowmass{\begin{center}\rule[-0.2in]{\hsize}{0.01in}\\\rule{\hsize}{0.01in}\\
\vskip 0.1in Submitted to the  Proceedings of the US Community Study\\ 
on the Future of Particle Physics (Snowmass 2021)\\ 
\rule{\hsize}{0.01in}\\\rule[+0.2in]{\hsize}{0.01in} \end{center}}

\begin{document}

\title{Snowmass 2021: Quantum Sensors for HEP Science - Interferometers, Mechanics, Traps, and Clocks}

\author{Oliver~Buchmueller}
\affiliation{Department of Physics, Blackett Laboratory, Imperial College, Prince Consort Road, London, SW7 2AZ, UK}
\author{Daniel Carney}
\affiliation{Lawrence Berkeley National Laboratory, Berkeley, CA, USA}
\author{Thomas Cecil}
\affiliation{Argonne National Laboratory, Lemont, IL, USA}
\author{John Ellis}
\affiliation{Physics Department, King's College London, Strand, London WC2R 2LS, UK}
\author{R.F. Garcia Ruiz}
\affiliation{Massachusetts Institute of Technology, Cambridge, MA, USA}
\author{Andrew A. Geraci}
\affiliation{Center for Fundamental Physics, Northwestern University, Evanston, IL, USA}
\author{David Hanneke}
\affiliation{Department of Physics \& Astronomy, Amherst College, Amherst, MA, USA}
\author{Jason Hogan}
\affiliation{Department of Physics, Stanford University, Stanford, CA, USA}
\author{Nicholas R. Hutzler}
\affiliation{Division of Physics, Mathematics, and Astronomy, California Institute of Technology, Pasadena, CA, USA}
\author{Andrew Jayich}
\affiliation{Department of Physics, University of California Santa Barbara, Santa Barbara, California, USA}
\author{Shimon Kolkowitz}
\affiliation{University of Wisconsin - Madison, Madison, WI, USA}
\author{Gavin W. Morley}
\affiliation{Department of Physics, University of Warwick, UK}
\author{Holger M\"uller}
\affiliation{Department of Physics, University of California, Berkeley, CA, USA}
\author{Zachary Pagel}
\affiliation{Department of Physics, University of California, Berkeley, CA, USA}
\author{Cristian Panda}
\affiliation{Department of Physics, University of California, Berkeley, CA, USA}
\author{Marianna S. Safronova}
\affiliation{Department of Physics and Astronomy, University of Delaware, Newark, DE, USA}


\snowmass

\maketitle

\begin{spacing}{.5}
\tableofcontents
\end{spacing}

\section{Executive summary}

A wide range of quantum sensing technologies are rapidly being integrated into the experimental portfolio of the high energy physics community. Here we focus on sensing with atomic interferometers; mechanical devices read out with optical or microwave fields; precision spectroscopic methods with atomic, nuclear, and molecular systems; and trapped atoms and ions. We give a variety of detection targets relevant to particle physics for which these systems are uniquely poised to contribute. This includes experiments at the precision frontier like measurements of the electron dipole moment and electromagnetic fine structure constant and searches for fifth forces and modifications of Newton's law of gravity at micron-to-millimeter scales. It also includes experiments relevant to the cosmic frontier, especially searches for gravitional waves and a wide variety of dark matter candidates spanning heavy, WIMP-scale, light, and ultra-light mass ranges. We emphasize here the need for more developments both in sensor technology and integration into the broader particle physics community.

\section{Introduction and background}

Progress in fundamental physics relies on performing measurements which can probe regions of some parameter space which have never been explored, in order to test whether our current models of particle physics need new ingredients. This requires the continuous development of ultra-sensitive detection systems. In recent years, numerous experimental programs have progressed to the point that the detectors are limited, or even enhanced, by the very laws of quantum mechanics. Such ``quantum sensors'' will continue to play an increasing role in the search for new fundamental physics. 

In this white paper, we focus on quantum sensors consisting of interferometers using atomic \cite{RevModPhys.81.1051} and mechanical \cite{RevModPhys.86.1391} systems, atomic \cite{RevModPhys.87.637} molecular and nuclear \cite{peik2003nuclear} clocks, and trapped neutral atoms \cite{ashkin1997optical} and ions \cite{RevModPhys.75.281}. These systems are used currently or in the near future in searches for gravitational waves \cite{aasi2015advanced,abe2021matter}, dark matter \cite{van2015search,carney2021mechanical}, ``fifth force'' and extra dimensional modifications to Newton's law \cite{hoyle2001submillimeter}, and dynamical sources of dark energy \cite{hamilton2015atom,sabulsky2019experiment}. They also provide our best measurements and set our current best limits on a number of crucial fundamental parameters, including the electromagnetic fine structure constant \cite{morel2020determination} and the electron's permanent electric dipole moment \cite{andreev2018improved}. 

The goal of this white paper is to provide an overview of the current status and future directions of these systems. In particular, we aim to outline the key open questions and technological requirements necessary to maximize the impact of these quantum sensors on particle and high energy physics. While we will highlight some central science targets motivating these developments, most of the motivation for specific particle physics models is relegated to white papers in the relevant Snowmass frontiers.

\section{Atom Interferometers}
\label{sec:AI}

Atom interferometry is a growing field with a variety of fundamental physics applications.  Science opportunities include gravitational wave detection~\cite{dimopoulos2008atomic,hogan2011atomic,Yu2011,graham2013new,canuel2018MIGA,Canuel2019ELGAR,kolkowitz2016gravitational,ZAIGA2020,abou2020aedge,Badurina_2020,Graham:2016plp, graham2017mid}, searches for ultralight (wave-like) dark matter candidates~\cite{arvanitaki2018search,Graham:2015ifn} and for dark energy~\cite{Hamilton2015}, tests of gravity and searches for new fundamental interactions (``fifth forces'')~\cite{Rosi2014,Biedermann2015,Rosi2017b,fray2004atomic,Schlippert2014ep,Zhou2015ep,Barrett2016,kuhn2014bose,barrett2015correlative,PhysRevLett.113.023005,PhysRevA.88.043615,Hartwig2015,asenbaum2020atom,williams2016quantum,berge2019exploring}, precise tests of the Standard Model~\cite{Bouchendira2011,parker2018measurement}, and tests of quantum mechanics~\cite{Arndt2014,Bassi2013,Nimmrichter2013,Bassi2017,altamirano2018gravity,kovachy2015quantum, asenbaum2016phase,xu2019probing,zych2011quantum,Roura2020}.  Such experiments take advantage of the ongoing evolution of the precision and accuracy of atomic sensors.  Optical lattice clocks now regularly attain 18 digits of frequency resolution~\cite{hinkley2013atomic,bloom2014optical} and beyond~\cite{Marti2018image,mcgrew2018atomic}, while atom interferometers continue to improve both in inertial sensing applications~\cite{bongs2019taking} and in precision metrology, including measurements of Newton's gravitational constant~\cite{fixler2007atom,PhysRevLett.100.050801,Rosi2014} and the fine structure constant~\cite{Bouchendira2011,parker2018measurement}, and testing the Equivalence Principle~\cite{fray2004atomic,Schlippert2014ep,Zhou2015ep,Barrett2016,kuhn2014bose,barrett2015correlative,PhysRevLett.113.023005,PhysRevA.88.043615,Hartwig2015,asenbaum2020atom,williams2016quantum,berge2019exploring}. The broad scientific potential of long-baseline quantum sensor networks has been widely recognized~\cite{WalsworthReport,PreskillReport,battaglieri2017us}. These sensors are noted for their potential use in searching for new fundamental forces, dark matter, and other dark sector ingredients~\cite{WalsworthReport}. 

There is widespread, growing international interest in pursuing long-baseline atomic sensors for gravitational wave detection~\cite{BertoldiGWOverview}. This has sparked a number of proposals for both space-based instruments and terrestrial detectors, some of which are already under construction today. In France, significant progress has been made towards the $200~\text{m}$ baseline underground gravitational wave detector prototype MIGA (Matter-wave laser based Interferometer Gravitation Antenna)~\cite{canuel2018MIGA}. A follow-on proposal has called for the construction of ELGAR (European Laboratory for Gravitation and Atom-interferometric Research)~\cite{Canuel2019ELGAR,Canuel2020ELGARtechnologies}, an underground detector with horizontal $32~\text{km}$ arms aiming to detect gravitational waves in the mid-band (infrasound) frequency range. In China, work has started to build ZAIGA (Zhaoshan long-baseline Atom Interferometer Gravitation Antenna)~\cite{ZAIGA2020}, a set of $300~\text{m}$ vertical shafts separated by kilometer-scale laser links that will use atomic clocks and atom interferometry for a wide range of research, including gravitational wave detection and tests of general relativity. In the UK, AION (Atom Interferometer Observatory and Network)~\cite{Badurina_2020} aims to progressively construct atom interferometers at the 10- and then 100-meter scale, in order to develop technologies for a full-scale kilometer-baseline instrument for both gravitational wave detection and dark matter searches.  In the US, MAGIS (Matter-wave Atomic Gradiometer Interferometric
Sensor) is a research program that is developing long-baseline atom interferometers based on narrow single-photon clock transitions for application to gravitational wave detection and searches for ultralight dark matter.  As the first step in this program, MAGIS-100 is a 100-meter-scale vertical atom interferometer currently under construction that will serve as a pathfinder for a future km-scale instrument.  A variety of space-based gravitational wave detectors have also been proposed to access the lower frequency ranges inaccessible to terrestrial observatories. These proposals are based both on optical lattice atomic clocks~\cite{kolkowitz2016gravitational,ebisuzaki2020ino} and atom interferometers~\cite{graham2017mid,hogan2016atom,abou2020aedge,hogan2011atomic,loriani2019atomic}, two technologies that are in fact closely related~\cite{norcia2017role}.

\subsection{Overview of the technique}

In light-pulse atom interferometry, laser pulses are used to coherently split, redirect, and recombine matter waves~\cite{Borde1989,kasevich1991atomic,tino2013atom,berman1997atom,Hogan2009}. The manipulation of matter waves is achieved through the stimulated absorption and emission of photons, driving transitions between two long-lived atomic states. Conventional atom interferometry uses alkali atoms like Rb and Cs, which make use of a pair of counter-propagating laser beams to drive two-photon Raman or Bragg transitions~\cite{muller2008, kasevich1991atomic}. A new variation of atom interferometry takes advantage of long-lived excited states in alkaline-earth-like atoms such as Sr~\cite{graham2013new}, which can be resonantly driven by a single laser beam and are used in some of the world's most precise atomic clocks~\cite{nicholson2015systematic,campbell2017fermi}.

In a gradiometer configuration, two identical atom interferometers are run simultaneously on opposite ends of a baseline, using the same laser sources. A comparison of the individual atom interferometer signals yields a differential measurement that enables the cancellation of noise common to both interferometers, such as the laser phase noise and vibration~\cite{mcguirk2002sensitive,Yu2011,graham2013new}. For atom interferometers using two-photon transitions driven by counter-propagating laser beams, laser frequency noise does not exactly cancel due to the asymmetry in the light travel times to the atoms~\cite{PhysRevD.78.042003,Yu2011,graham2013new}. However, in atom interferometers based on single-photon clock transitions, the laser pulses are derived from a single laser and both interferometers are driven by nominally identical laser pulses.  This in principle enables superior common-mode rejection of noise, allowing for the possibility of, for example, gravitational wave detection using a single baseline.

In an atom interferometer gravitational wave detection, freely-falling atoms on each end of a long baseline simultaneously acting as inertial references and as precise clocks.  Laser light propagates between the two atom ensembles, driving transitions between the atomic states and encoding the light travel time across the baseline onto the phase of the atomic wavefunction.  As a result, the differential phase measurement between the two atom interferometers is sensitive to variations in both the baseline and the atomic energy level splitting. A passing gravitational wave modulates the baseline length, while coupling to an ultralight dark matter field can cause a modulation in the energy levels. This combines the prospects for both gravitational wave detection and dark matter searches into a single detector design, and both science signals are measured concurrently.

\subsection{Science with tabletop atom interferometers}

\begin{figure}
    \centering
    \includegraphics[width=4.5in]{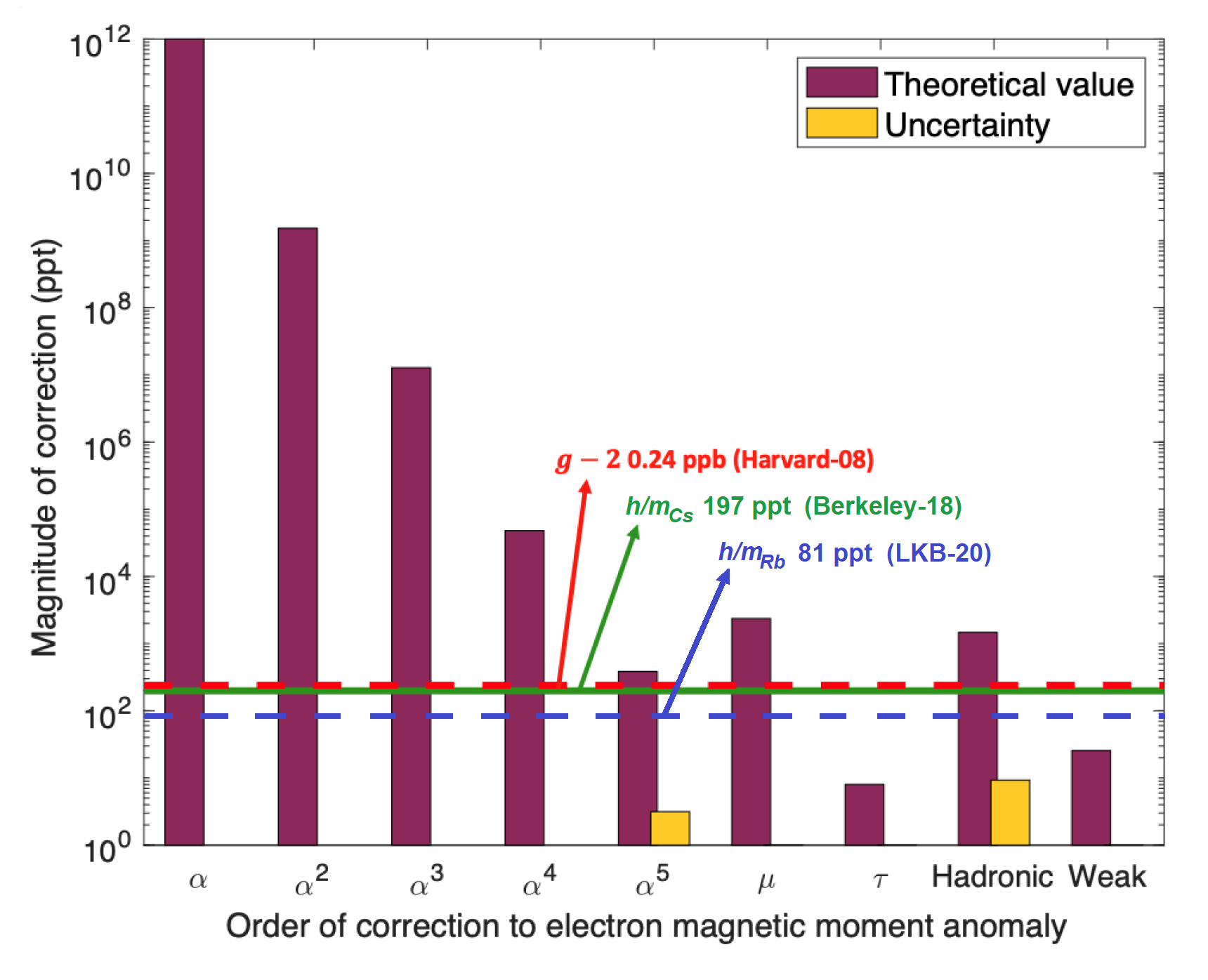}
    \caption{A breakdown of the Standard Model terms being tested by current measurements of $\alpha$ and $g_e-2$. Current measurements are probing 5th order QED, 1st order muonic contributions, and hadronic contributions. Near-term future experiments will begin to resolve weak force contributions.
    }
    \label{fig:SMcontributions}
\end{figure}

\textbf{The fine-structure constant, anomalous magentic moments, and testing the Standard Model.}
Atom interferometry is a powerful tool for measuring $h/m_{\rm At}$, where $h$ is the Planck constant and $m_{\rm At}$ the mass of an atom, and (by combination with the Rydberg constant and the mass ratio between the electron and the atom) the fine structure constant $\alpha$, a central parameter when testing quantum electrodynamics (QED) and the Standard Model. The most precise test of QED uses atom interferometry measurements of $\alpha$ as input for the standard-model prediction for the electron's anomalous magnetic moment $a_e$, and compares this result to an experimental measurement \cite{alpha_berkeley, alpha_lkb, g-2_e}. With accuracies currently at $\sim 10^{-10}$, this allows testing 5th-order QED contributions, the contribution of the muon, and tests the hadronic contribution at a part in 10 (see Fig. \ref{fig:SMcontributions}). 

\begin{figure}
    \centering
    \includegraphics[width=4.5in]{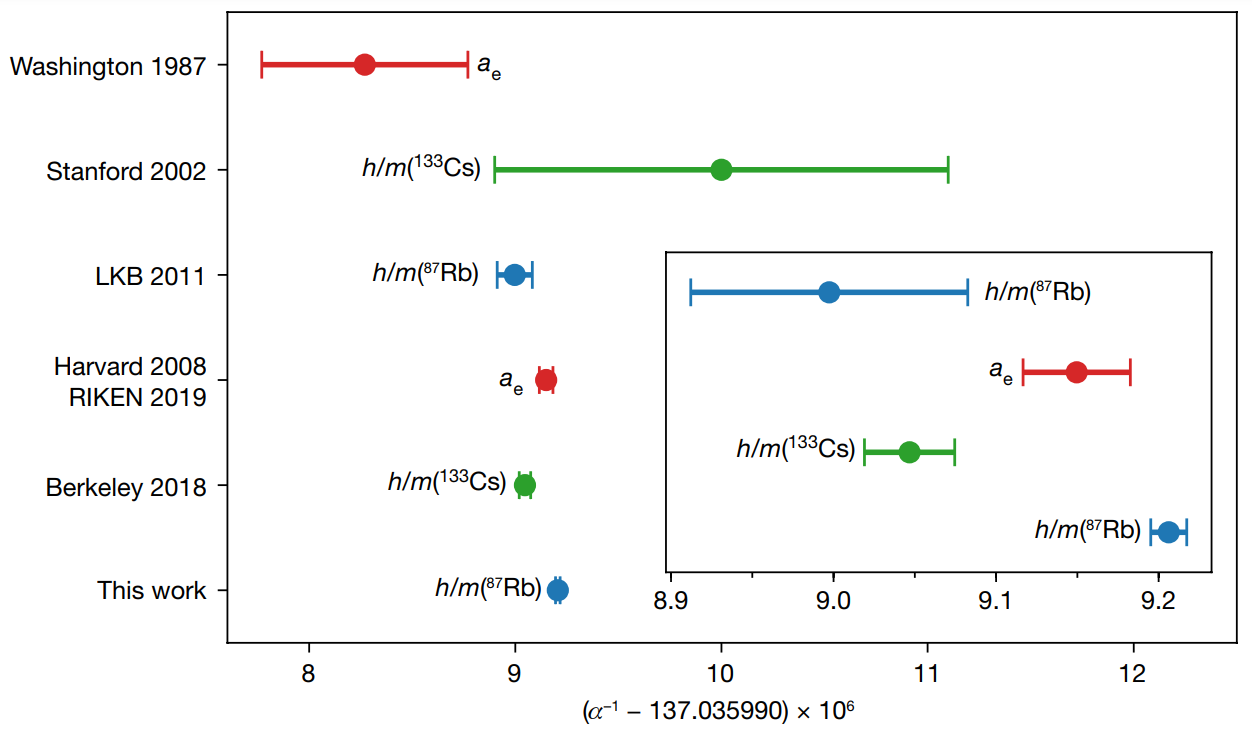}
    \caption{Comparison of the most recent measurements of $\alpha$ and of $g_e-2$, with $1\sigma$ error bars shown. Green and blue data points correspond to atom interferometry measurements of $\alpha$ using Cesium and Rubidium respectively. Red data points are calculated from measurements of $g_e-2$ using the Standard Model prediction. This figure was taken from \cite{alpha_lkb}.
    }
    \label{currentAlphaMeasurements}
\end{figure}

The motivation for improved measurements of the fine structure constant is strong: A tenfold improvement in accuracy would, e.g., resolve weak force contributions and would constitute a sensitive and broad search for dark constituents of the vacuum, such as the dark photon, in a parameter space that is complementary to other searches that are currently underway or proposed. In addition, while there is only a slight tension between each of the two most recent determinations of $\alpha_s$ and the anomalous magnetic moment, there is a strong tension between themselves, as they deviate from the standard-model prediction in opposite directions (Fig. \ref{currentAlphaMeasurements}) \cite{alpha_berkeley, alpha_lkb}. It is likely that only an improved measurement will be able to resolve this tension. 
 
Added relevance of measuring $\alpha$ derives from Fermilab's highly anticipated measurement of the muon anomalous magnetic moment $a_\mu$, which has reproduced the long-standing tension with the Standard Model \cite{g-2_mu}.  From naive scaling by the square of the mass ratio $m_e/m_\mu$, the 4.2-$\sigma$ deviation of $[(a_\mu)_{\rm exp}-(a_\mu)_{\rm thy}]/a_\mu \sim (2.2\pm 0.5) \times 10^{-6}$ implies a deviation of $[(a_e)_{\rm exp}-(a_e)_{\rm thy}]/a_e  \sim 5.0 \times 10^{-11}$, which is just a bit smaller than what can be resolved at the current experimental accuracy. Thus, improved measurements of $\alpha$ and $a_e$ are highly likely to give valuable input to understanding the unexplained anomaly seen for the muon. 

An improved measurement of the fine-structure constant that is currently underway at Berkeley has a target sensitivity of $2\times 10^{-11}$, a 10-fold improvement over Berkeley's previous measurement and a 4-fold improvement over the most recent one from LKB \cite{alpha_lkb}. In the long run, this accuracy could be further improved. The experiment is designed to strongly reduce systematic effects from large-scale and small-scale curvature in the laser-beam wavefronts, while also improving contrast and overall sensitivity. The experiment is already partially functional; atoms are being launched in an atomic fountain, and the interferometry laser system is being finalized. It is expected to perform interferometry and begin characterizing the error budget soon.

\textbf{Dark energy and tests of quantum gravity by atom interferometry in an optical cavity.} Atom interferometers have recently taken advantage of optical cavities, which provide power enhancement, spatial mode filtering and clean optical wavefronts. We have used an atom interferometer inside an optical cavity to place stringent limits on a wide class of dark energy candidates \cite{Hamilton2015,Jaffe2017,Sabulsky2019}, including the chameleon and other theories that reproduce the observed cosmic acceleration. Such theories predict that dark energy consists of a light scalar field, which might be detectable as a “fifth force” between normal-matter objects, in potential conflict with precision tests of gravity. Chameleon fields and other theories with screening mechanisms, can evade these tests by suppressing the forces in regions of high density, such as the laboratory. Using a cesium matter-wave interferometer near a small source mass in an ultrahigh-vacuum chamber, we reduced the screening mechanism by probing the field with individual atoms rather than with bulk matter.

\begin{figure}[t]
    \centering
    \includegraphics[width=4.5in]{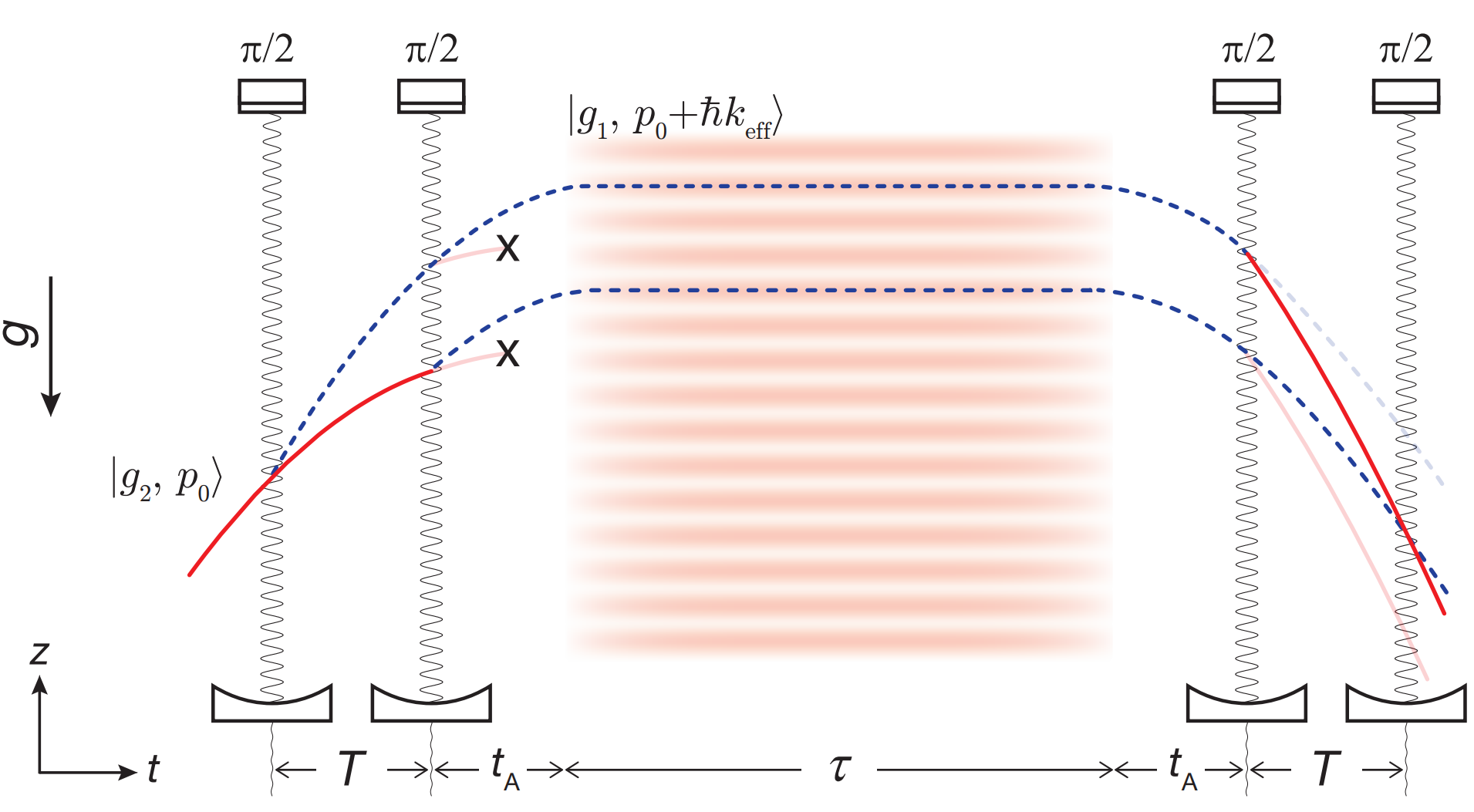}
    \caption{ Atom trajectories in the lattice interferometer. The red solid lines and blue dashed lines show trajectories for the two hyperfine states of the Caesium atom, $F = 3$ and $F = 4,$ respectively. Each pulse pair is separated by a time $T$. Between the $\pi$/2 pulses and the lattice hold, atoms move in free fall towards the apex of their trajectory for a time $t_A$. At the apex, atoms are loaded in a far-detuned optical lattice formed by the mode an optical cavity (red stripes) and remain held in the lattice for a time $\tau$. (Adapted from Ref \cite{Xu2019}.)
    }
    \label{LatticeInt}
\end{figure}

More recently, clean wavefronts of the optical cavity have been used to hold atoms against gravity (Fig. \ref{LatticeInt}), realizing an interrogation time of 20 seconds (and, in unpublished work, even 25 seconds) by suspending the spatially separated atomic wave packets in an optical lattice \cite{Xu2019}. This surpassed the limited time available to interferometers with freely falling atoms by more than an order of magnitude. This record coherence is enabled by the smooth lattice wave fronts, which are mode-filtered by the optical cavity. The trapped geometry also suppresses phase variance due to vibrations by three to four orders of magnitude, overcoming a dominant noise source in atom-interferometric gravimeters. Recent efforts have been focused on diagnosing sources of decoherence, including investigations of various imperfections of the optical trapping potential or optical beamsplitter pulses, increases in the lattice lifetime and reduction of noise.

Trapped interferometers are expected to enable new experiemtns in a wide range of fields, from metrology (e.g., portable, gravity sensors that are insensitive to vibration and tilt) to fundamental applications. The record coherence time and ability to hold atoms at locations that are maximally sensitive to the interaction with small source masses will enable searches for fifth forces with increased precision, as well as measurement of a phase shift due to a difference in the gravitational potential in the absence of forces, through the gravitational Aharanov-Bohm effect \cite{Hohensee2012}. This technology will also enable a recently proposed test of the ability of the gravitational field to generate entanglement between the atomic interferometer and a mechanical oscillator, which requires exquisite coherence time and sensitivity \cite{Carney2021}.


\subsection{Science with long baseline atomic sensors}

\textbf{Gravitational waves.} Atomic sensors are promising for filling the frequency gap between LIGO/Virgo/KAGRA and LISA, observing gravitational waves in the mid-band, roughly $30~\text{mHz}$ to $3~\text{Hz}$ \cite{abe2021matter}.  The mid-band may be optimal for observing signals of cosmological origin. This frequency range is above the white dwarf ``confusion noise'' but can still extend low enough in frequency to see certain cosmological sources, including certain models of inflation~\cite{Graham:2016plp}.   Furthermore, thermal phase transitions in the early universe at scales above the weak scale~\cite{randall2007gravitational,Caprini:2009fx,Caprini:2009yp,Konstandin:2011dr,Schwaller_2015,Hindmarsh:2017gnf,Cutting:2019zws,Breitbach_2019,Caprini:2019egz}, or quantum tunnelling transitions in cold hidden sectors~\cite{GarciaGarcia:2016xgv}, or networks of cosmic strings~\cite{depies2007stochastic}, or collapsing domain walls~\cite{Saikawa2017}, or axion dynamics in the early universe~\cite{Machado_2019,Machado_2020}, may produce detectable gravitational wave signals in this band.

There are also important known astrophysical sources in the mid-band.  Many black hole or neutron star binaries that are observed in the mid-band can later be observed by LIGO/Virgo/KAGRA once they evolve to higher frequencies.  Such joint observation would be a powerful new source of information. For example, measurements of early infall stages by an atomic sensor in the mid-band could give a prediction of the time and location of a LIGO/Virgo/KAGRA merger event, facilitating multi-messenger astronomy. In addition, a mid-frequency detector could observe mergers of intermediate-mass black holes that could not be observed by higher-frequency detectors, providing sensitive probes of modifications of gravity due to a graviton mass or Lorentz violation~\cite{Ellis:2020lxl}. Since the lifetimes of many sources in the mid-frequency band are comparable to the orbital period of the Earth, the mid-band is ideal for sky localization and prediction of merger events, and in fact they can be localized even by a single-baseline detector~\cite{graham2018localizing}. Thus, observations in the mid-band have the potential to be a powerful complement to detection by LIGO, and can significantly enhance multi-messenger astronomy.

In addition to being interesting and important astrophysically, observing compact objects such as black holes, neutron stars, and white dwarfs may well also teach us about particle physics.  For example, supernovae and other such extreme astrophysical objects have already been used to set some of the best limits on axions and other light particles, and  gravitational wave observations may allow observations of gravitational memory effects~\cite{Badurina:2021rgt} as well as more tests for new physics.  As just one example, superradiance around black holes~\cite{Arvanitaki:2009fg, Arvanitaki:2010sy,Brito:2015oca} may allow us to constrain or even discover such particles with future gravitational wave observations.

\textbf{Dark matter and new forces.} There are several strategies that can be employed to search for dark matter using atom interferometry.  First, dark matter that affects fundamental constants, such as the electron mass or the fine structure constant, will change the energy levels of the quantum states used in the interferometer, causing them to oscillate at the Compton frequency of the candidate dark matter particle.  This effect can be searched for by comparing two simultaneous atom interferometers separated along the baseline.  Second, dark matter that causes accelerations can be searched for by comparing the accelerometer signals from two simultaneous atom interferometers run with different isotopes (${}^{88}$Sr and ${}^{87}$Sr for example)~\cite{Graham:2015ifn}.  This requires running a dual-species atom interferometer, which is well established~\cite{PhysRevLett.113.023005,PhysRevA.88.043615,kuhn2014bose,overstreet2018effective}.  Third, dark matter that causes precession of nuclear spins, such as general axions, can be searched for by comparing simultaneous, co-located interferometers using Sr atoms in quantum states with differing nuclear spins.  See~\cite{Graham:2017ivz} for a discussion and potential sensitivities.

In addition to these dark matter searches, new fundamental particles may also be discovered by searching for new forces~\cite{WalsworthReport}. Ultralight particles that have highly suppressed interactions with Standard Model particles, often dubbed ``dark sectors'', emerge in a variety of beyond-the-Standard-Model frameworks. These theories include forces mediated by particles that can dynamically solve naturalness problems in the Standard Model, such as the strong CP problem (QCD axion~\cite{Moody:1984ba}) and the hierarchy problem (relaxion~\cite{Graham:2015cka}). Such forces can also arise in theories with extra dimensions~\cite{ArkaniHamed:1998nn} as well as supersymmetry~\cite{Dimopoulos:1996kp}.  These ultra-weak forces can be sourced either by a test mass or by the Earth itself.  While several of these particles are also ultralight dark matter candidates, there are alternative ways to search for the presence of these new fields, without necessarily requiring them to be dark matter. In principle there are two ways to do this. First, if the range of the new force is short, it can be observed by modulating the distance between a test mass and the atomic sensor. Second, long range forces sourced by the Earth other than gravity may lead to differential free-fall accelerations between different elements/isotopes. A comparison between atomic sensors made out of different atomic species could reveal the existence of such forces.

\subsection{Detector development}


The MAGIS concept~\cite{graham2013new,graham2017mid} takes advantage of features of both clocks and atom interferometers to allow for a single-baseline gravitational wave detector~\cite{PhysRevD.78.042003,Yu2011,graham2013new}.  It aims to detect gravitational waves in the scientifically rich, so-far unexplored `mid-band' frequency range between $0.01~\text{Hz}$ and $3~\text{Hz}$. This band lies below the sensitivity range of existing terrestrial interferometers (LIGO/Virgo) and above the frequency band of the planned LISA satellite detector. Simultaneously, the MAGIS concept enables the exploration of new regions of dark-sector parameter-space~\cite{BRNreport} by being sensitive to proposed scalar- and vector-coupled dark matter candidates in the ultralight range ($10^{-15}$~eV -- $10^{-14}$~eV).

MAGIS-100 is the first detector facility in a family of proposed experiments based on the MAGIS concept. The instrument features a 100-meter vertical baseline and is now under construction at the Fermi National Accelerator Laboratory (Fermilab). State-of-the-art atom interferometers are currently operating at the 10-meter scale~\cite{Dickerson2013,kovachy2015quantum,asenbaum2016phase,asenbaum2020atom,Hartwig2015,zhou2011development}, while a kilometer-scale detector is likely required to detect gravitational waves from known sources.  MAGIS-100 is thus a pathfinder for a future full-scale gravitational wave detector.  Terrestrial detectors with several kilometer vertical baselines are being studied.  Newtonian gravity gradient noise (GGN) is an important source of background noise for any terrestrial gravitational wave detector, including MAGIS~\cite{HarmsGGN}. GGN likely imposes a practical limit on the sensitivity at low frequencies for Earth-based gravitational wave detectors, and is one of the primary motivations for space-based detectors.  One possible mitigation that will be studied by MAGIS-100 is to use more than two atomic test masses along the baseline to attempt to distinguish GGN from a gravitational wave~\cite{PhysRevD.93.021101}.


The Atom Interferometric Observatory and Network (AION) project~\cite{Badurina_2020} envisages a staged Atom Interferometry programme, starting with a 10~m device and progressing via a 100~m experiment to a 1~km instrument. AION will enable exploration of the properties of ultra-light dark matter (DM) and gravitational waves (GWs) from the very early Universe and astrophysical sources in the mid-frequency band ranging from several mHz to a few Hz, intermediate between the sensitive ranges of LIGO/Virgo/KAGRA and LISA. The AION science programme spans a wide range of fundamental physics, astrophysics and cosmology, and aligns fully with the top priorities of international communities. It has been approved by the UK STFC with initial funding of about £10M. 

The ultimate sensitivity of the AION program will be reached by interoperating and networking with other  instruments around the world, similar to the existing LIGO-Virgo network, which will provide science opportunities not accessible to single detectors. AION has a close collaboration with MAGIS~\cite{MAGISLOI, MAGIS100, Graham:2017pmn} in the US, which also targets an eventual km-scale atom interferometer. AION and MAGIS complement several other terrestrial cold atom experiments that are currently  being prepared, such as MIGA~\cite{Canuel:2017rrp} and ZAIGA~\cite{Zhan:2019quq}, or being proposed, such as ELGAR~\cite{Canuel:2019abg}. The AION consortium contributes strongly to the design study of a mission proposal for an Atomic Experiment for Dark Matter and Gravity Exploration in Space (AEDGE)~\cite{abou2020aedge}, which was selected by European Space Agency (ESA) for presentation at its Voyage 2050 workshop. AEDGE would make direct use of AION and MAGIS technology.

Another satellite proposal, STE-QUEST~\cite{Battelier2021}, would use ultracold  atoms  in  quantum  superposition  states to test the  equivalence  principle with a sensitivity about three orders of magnitude beyond the best existing result obtained by the MICROSCOPE space mission in 2017~\cite{Touboul2017}.  Additional science goals are searches for different types of dark matter and tests of the foundations of quantum mechanics. The STE-QUEST Phase-1 proposal is currently under evaluation by ESA.


\section{Optomechanical sensors}

Mechanical sensors read out by optical \cite{RevModPhys.86.1391} or microwave \cite{blencowe2004quantum} light have enjoyed rapid development in the last few decades, leading to orders of magnitude improvements in sensitivity. Much of this development has come in tandem with LIGO \cite{aasi2015advanced}, which serves as an excellent example of the general concept of optomechanics. In particular, mechanical sensors are now routinely operated in the quantum regime, in which their sensitivity becomes dominated by quantum noise in the mechanics and/or readout system \cite{caves1981quantum}. Mechanical devices are uniquely suited to looking for signals which act coherently over a length scale around the size of the mechanical system, since the signal is coherently integrated into one or a few collective degrees of freedom, for example the center-of-mass motion.

A vast array of architectures are available. The mechanical elements can be as large as tens of kilograms \cite{aasi2015advanced} and as small as single ions or even electrons \cite{ivanov2016high}, and can also include collectively quantized degrees of freedom like phonons in both solids \cite{balram2016coherent,chu2017quantum} and liquids \cite{bahl2013brillouin,kashkanova2017superfluid}. They can be operated at frequencies anywhere below roughly GHz scales. Examples include mechanically suspended reflective pendula \cite{matsumoto2019demonstration}; optically levitated dielectrics \cite{yin2013optomechanics}, cold atoms \cite{brennecke2008cavity,brooks2012non}, and ions \cite{rohde2001sympathetic}; clamped nanomechanical membranes \cite{teufel2011sideband}; and magnetically levitated systems \cite{moody2002three,prat2017ultrasensitive}. Readout can be performed using both free-space and cavity optics, as well as microwave cavity/circuit QED systems.

In addition to their use in gravitational wave detection and precision measurements in metrology, optomechanical devices are rapidly being incorporated into the portfolio of detector systems useful for a number of high energy and particle physics targets. Building on classical proposals for neutrino and dark matter detection with nanoscale targets \cite{smith1991prospects}, proposals now exist to use optomechanical sensors for detection of ultra-light \cite{Graham:2015ifn,Arvanitaki:2015iga,Carney:2019cio,Manley:2019vxy,Manley:2020mjq}, MeV-to-TeV scale \cite{Carney:2021irt,Afek:2021vjy}, and ultra-heavy \cite{Carney:2019pza} dark matter (see \cite{carney2021mechanical} for an overview); neutrinos \cite{Domcke:2017aqj}; high-frequency gravitational waves \cite{Aggarwal:2020umq,Aggarwal:2020olq}; fifth-force modifications to Newton's law at tabletop scales \cite{Montoya:2021dqz}; deviations from standard quantum mechanics \cite{Gasbarri:2021sdm} (including ideas about gravitational breakdown of quantum mechanics \cite{Marshall:2002exi}); and tests of quantum properties of the gravitational interaction \cite{Kafri:2014hlh,Bose:2017nin,Carney:2018ofe}.

Numerous experiments are already ongoing or coming online in the near future. Here we discuss a few examples of methods which have already achieved significant results and/or show substantial promise.

\textbf{Torsion balances.} Sensitive torsion balance experiments such as those performed by the Eotvos-Washington group are a proven method to search for both static \cite{EPtests1,wagnerEPV} and time-varying equivalence principle (EP) violating interactions, novel spin-dependent forces \cite{Terrano:2015}, and tests of non-Newtonian gravity \cite{kapner2007,Lee:2020zjt}. They are one of the most promising paths forward for these studies as their sensitivity increases and the understanding of background systematic errors due to patch charges and other surface forces improves. 

\textbf{Opto-mechanical interferometers.}
Interferometric cavity-based sensors have attained remarkable sensitivity, allowing the detection of gravitational waves from from inspiraling black hole mergers resulting in strain at the level of one part in a billion-trillion \cite{LIGOfirst}. Quantum-based squeezing techniques could enhance the sensitivity of opto-mechanical systems by an order of magnitude or more. Opto-mechanical systems are also detectors which can be used to search for wave-like dark matter, as many models predict a strain-like behavior in objects as a result of the oscillating background dark matter bosonic wave.  For example, meter-scale optical cavities can be used to search for dilatonic dark matter in the audio band. The comparison of a cavity with suspended mirrors against a cavity whose length is ``fixed'' by a rigid spacer is sensitive to oscillations in both $\alpha$ and $m_e$ \cite{Stadnik:2016DM-cavity,Geraci:2019DM-cavity} that could result in a measurable strain.
 Gravitational wave interferometer detectors have also been used to constrain ultralight DM \cite{ligodarkphotons2021,Grote:2019DM-LIFO}. 

\textbf{Resonant mass detectors.}
Weber-bar-type detectors with longitudinal acoustic modes enabling laboratory-scale detectors to probe for scalar UDM below the E\"{o}t-Wash torsion constraints \cite{Arvanitaki2016_SoundDM,manley2020searching}. AURIGA is a resonant mass GW detector that has searched for scalar ultra-light dark matter and set some of the strongest constraints at $\sim 1$ kHz \cite{Arvanitaki2016_SoundDM,AURIGA}. In the DAMNED experiment \cite{Savalle:2021DAMNED}, an ultrastable optical cavity behaves effectively as a multimode resonant mass detector, where broadband readout is accomplished with an optical fiber interferometer. These approaches are powerful methods for searching in the vacinity of their resonant frequencies.

\textbf{Nanomechanical systems.}
Complementing the atom interferometry-based approaches discussed above, a number of groups are working on optomechanical accelerometer searches for ultra-light dark matter.  Vector dark matter for example coupled to Baryon-Lepton number (B-L) can produce material-dependent differential acceleration between two bodies, which would take the form $a_{\rm diff}(t) \approx g_{B-L} \Delta a_0 \sin(m_{\phi} t)$, where $\Delta = Z_1/A_1 - Z_2/A_2$ is the difference in the neutron-nucleon ratio of the bodies and $m_{\phi}$ is the DM mass.  Resonant enhancement of this deformation can be achieved by having two masses made of different materials are bound by a spring. The two masses may be fashioned into mirrors, forming an optical cavity for displacement based readout \cite{manley2021searching}.   Cavity optomechanical systems offer a diversity of platforms for vector ultra-light Dark matter detection based on the heterogeneous dimer model.  For example, Ref. \cite{manley2021searching} proposed a specific design based on a membrane optomechanical system that features ultra-high mechanical $Q$, frequency tuning via stress, quantum-limited cavity-based readout, and cryogenics.  In the past decade, it has become possible to prepare quantum states of motion of nano-/micro-scale solid-state mechanical oscillators \cite{cleland2010}. 
Very recently, a kilogram-scale mechanical oscillators can be prepared close to their motional quantum ground state through measurement-based feedback control \cite{10kggroundstate}. Such an experiment with a pair of
milligram-scale masses can offer a potential route towards a non-interferometric test of the role of gravity in quantum entanglement.

\textbf{Levitated particles.}
In high vacuum, optically-levitated dielectric nanospheres achieve excellent decoupling from their environment, making ultra-precise force sensing \cite{Ranjit:2016} and acceleration \cite{Geraci:matter_wave, Moore2017,novotnydrop} sensing achievable.  There are several applications of these sensors for tests new physics \cite{GeraciMoore2020}, including testing the Newtonian gravitational inverse square law at micron length scales \cite{Geraci2010}, searching for gravitational waves \cite{GeraciGravityWave}, searching for millicharged particles \cite{Afek:2020}, and searching for dark matter. For example, in searches for dark matter, the authors in Ref. \cite{Monteiro:2020wcb} performed a search for novel, GeV-TeV particle dark matter candidates with a levitated, $\mu$g-scale dielectric sphere read out optically. 
Advances in sensitivity made possible by pushing the sensitivity of these sensors into the quantum regime along with improved understanding and mitigation of systematic effects due to background electromagnetic interactions will enable several orders of magnitude of improvement in the search for new physics beyond that Standard model. 

By developing new methods based on interferometry with levitated nanoparticles, experimental proposals have been presented for using macroscopic superpositions of levitated nanoparticles to test whether the gravitational field can entangle the states of two masses \cite{Marletto:2017,Bose:2017nin},  
e.g. where embedded spins in the masses can be used as a witness to probe the entanglement \cite{Bose:2017nin}. The spin in each mass could be a nitrogen-vacancy center (NVC) in diamond. By putting the NVC into a spin superposition and applying an inhomogeneous magnetic field, a superposition of forces is present on the diamond \cite{ScalaPRL2013, YinPRA2013, WanPRA2016}. This could produce a spatial superposition, but to make it larger it would be beneficial to drop the nanodiamond as the trapping forces tend to oppose the forces creating the spatial superposition \cite{WanPRL2016}. To further increase the spatial superposition distance, motional dynamic decoupling would be used \cite{PedernalesPRL2020}, and to protect the NVC spin coherence, spin dynamic decoupling should be used in addition, which could be done by dropping the nanodiamond past magnetic teeth \cite{WoodPRA2022}. The use of a Casimir screen could help to prevent the Casimir effect from entangling the two nanodiamonds \cite{vandeKampPRA2020}. Performing the experiment in a platform inside a drop tower may be needed to reduce the relative acceleration noise \cite{TorosPRR2021}. An extension of these ideas could be used to build a compact gravitational wave detector \cite{MarshmanNJP2020}. 

Such experiments require an ultra-high-vacuum ultra-low-vibration cryogenic environment to minimize spurious environmental perturbations and technical noise. Along the way towards testing the quantum nature of gravity, this research program would strongly test proposals that macroscopic superpositions can cause objective collapse of the wavefunction \cite{BassiRMP}. One experiment that has been demonstrated recently towards this work is a full-loop coherent Stern-Gerlach interferometer with single atoms \cite{MargalitSA21}. 


Moving forward, a number of key opportunities exist to increase the utility of these devices in the search for new physics. A critical one is the need for new theoretical ideas about potential new signals! At the level of the detector technologies, an important frontier is in advanced quantum techniques to get sensitivities at and beyond the so-called Standard Quantum Limit (SQL) \cite{caves1981quantum}. The most common, well-demonstrated method to go beyond the SQL is the use of squeezed light \cite{aasi2013enhanced,Magrini:2022rlc,Militaru:2022pen}. Another, less-studied option is the use of backaction evasion techniques \cite{chen2011qnd}. Further theoretical development and implementation of these techniques in disparate situations and physical architectures, especially in broadband sensing problems, will be of crucial importance in the next decade. Leveraging multiple sensors (``networks'') and entanglement between them can similarly enable detection beyond the SQL; using these ideas in searches for new physics would be extremely interesting.


\section{Atomic, nuclear, and molecular clocks \& precision spectroscopy}
\label{clocks}

Optical clock precision has improved by more than three orders of magnitude in the past fifteen years \cite{ludlow_optical_2015}, enabling tests of the constancy of the fundamental constants and local position invariance \cite{2021YbclockAlpha}, dark matter searches \cite{2020SrcavityDM}, tests of the Lorentz invariance \cite{2019YbLLI}, and tests of general relativity \cite{2020Skytree}. Future clock development will provide further orders of magnitude of improvement for these experiments. Deployment of high-precision clocks in space will also open the door to new applications, including precision tests of gravity and relativity \cite{2022FOCOS}, searches for a dark-matter halo bound to the Sun \cite{2020SpaceQ}, and gravitational wave detection in wavelength ranges inaccessible on Earth \cite{2016clockGW,2022clockGW}.

These applications motivate the development of novel clocks with enhanced sensitivity to variation of fundamental constants and, therefore, dark matter.
If  the fundamental constants
are space-time dependent, so are atomic and nuclear spectra including clock transition frequencies. Variations of fundamental constants due to interactions with dark matter will change the rate at which an atomic clock ticks, with the magnitude of the effect dependent on the nature of the specific clock transition, as the frequencies of different clock transitions depend differently on fundamental constants.
Ultralight dark matter can source the oscillatory and transient variation of fundamental constants that can be detected by comparing frequencies of two different atomic clock transitions, a clock and a cavity \cite{Safronova2018,2020SrcavityDM}, or otherwise identical spatially separated atomic clocks \cite{roberts2017search}.

Dark matter detection with clocks is described in detail in CF2 Snowmass paper  ``New Horizons: Scalar and Vector  Ultralight Dark Matter'' as we focus on the proposed R\&D developments here.

 All current atomic clocks are based on (1) transitions between the hyperfine substates of the atomic ground state (microwave clocks)  or (2) transitions between different electronic levels (optical clocks) \cite{ludlow_optical_2015}. The frequency ratio of two optical clock frequency is only sensitive to the variation of $\alpha$ and optical atomic clocks can probe the $\phi F^{\mu \nu}F_{\mu \nu}$ SM-DM coupling and the corresponding quadratic coupling.
The degree of sensitivity to  the variation of $\alpha$ tends to increase for states with similar electronic configurations for atoms with heavier nuclei;  details of the electronic structure can lead to significantly larger enhancement factors. Among all presently operating clocks Yb$^+$ clock based on the octupole transition has the largest (in magnitude) sensitivity factor $K=-6$ \cite{FlaDzu09}. Therefore, it is essential to develop new clocks with much larger sensitivity factors.

The most recent limits on the drift of the fine-structure constant come from
the comparison of two optical clocks based on the electric quadrupole (E2) and the  electric octupole (E3) transition of $^{171}$Yb$^+$ \cite{2021YbclockAlpha}. A comparison of the $^{171}$Yb$^+$ octupole (E3) transition to the Cs hyperfine microwave clock provides the best limit on the drift of the ratio $m_p/m_e$. Repeated measurements  over several years are analysed for potential violations of local position invariance. 

The dark matter limits from the atomic clocks and precision spectroscopy come from re-analyses of $\alpha$ drift data for Dy/Dy \cite{PhysRevLett.115.011802}, Rb/Cs microwave clocks~\cite{PhysRevLett.117.061301}, and  Al$^+$/Hg$^+$ optical clocks~\cite{Beloy2021}. New experiments included clock-comparison experiments with Yb/Al$^+$ and Yb/Sr  clock pairs limits~\cite{Beloy2021} and comparison of hydrogen maser and strontium optical clock with a cryogenic crystalline silicon cavity (H/Si and Sr/Si) ~\cite{Kennedy:2020DM_atom-cavity}.

\subsection{Clock R\&D}

Several R\&D directions are being pursued to drastically improve the reach of clock experiments for DM detection. These experimental efforts are complimented by the development of high-precision atomic theory~\cite{Ir} and particle physics model building~\cite{Graham:2015cka, Flacke:2016szy,Banerjee:2020kww}.

\textbf{ Development of clock networks at the new level of precision~\cite{roberts_search_2020,2022QSNET}} --
Many of the emerging applications of optical atomic clocks as particle physics detectors require comparisons between clocks in different spatial locations, such as tests of local position invariance through gravitational redshift measurements \cite{will2014confrontation,2020Skytree}, searches for dark matter through spatial and temporal variations in clock transition frequencies \cite{roberts_search_2020,2022QSNET}, and gravitational wave detection across a long baseline (see Sec.~\ref{sec:ClockGW}). In addition, optical clock frequency ratio measurements, which can be used to search for dark matter or other sources of variations in fundamental constants, require comparisons between clocks with different atomic species, which are often located at different standards institutes around the world. This motivates the development of larger and more integrated clock networks, and also the development of portable atomic clocks that can be transported to different nodes of a network or placed in more extreme environments and distant locations, including in space.

There has been considerable recent progress on the demonstration and development of clock networks at new levels of precision, including recent clock comparisons at new levels of precision between multiple species over free-space and fiber-based links \cite{Beloy2021}, with portable optical clocks \cite{grotti2018geodesy,2020Skytree}, and across large fiber networks\cite{roberts_search_2020,2022QSNET}. Ultimately high performance space-based optical clocks will be required in order to connect clock networks across continents and to extend the spatial extent of clock networks beyond the limits imposed by the size of the Earth. Such space-based clocks would also enable new tests of fundamental physics \cite{derevianko2021fundamental}. This motivates research and development into improving the size, weight, and power (SWaP) requirements for optical clocks and into making hardier and more robust clocks that can operate autonomously for extended periods of time \cite{gellesch2020transportable,grotti2018geodesy,2020Skytree,hannig2019towards}. While progress has been made, optical atomic clocks currently remain laboratory-scale experiments that require regular maintenance and human intervention.

\textbf{Improvements in clock performance and prospects for the use of quantum entanglement to measure beyond the standard quantum limit~\cite{2020PietSqueezing}} --
Both the stability and uncertainty of optical clocks are expected to continue to rapidly improve.
Recently, Ref.~\citep{2022JunSr} demonstrated atomic coherence of 37 s on an optical transition in a differential comparison and an expected
single clock stability of $3.1 \times 10^{-18}$ at 1 s using macroscopic samples \cite{2022JunSr}. However, the record stability for independent clock comparisons remains more than an order of magnitude above this at $4.7 \times 10^{-17}$ at 1 s \cite{oelker2019demonstration}, leaving considerable work to be done to reach the levels of performance promised by differential comparisons. The stability of the local oscillator used to probe each clock is of critical importance for independent clock comparisons, as it limits coherent interrogation times and often prevent the clocks from reach the quantum projection noise limit through the Dick effect \cite{westergaard2010minimizing, schioppo2017ultrastable}. Remarkable progress has already been made, including the demonstration of a linewidth of $8~\textrm{mHz}$ and stability of $4\times10^{-17}$ at 1 s for a single crystal silicon cavity \cite{matei20171}, making it a viable timescale itself \cite{milner2019demonstration}. Nevertheless, further improvements are necessary, and can potentially be achieved through the integration of multiple ultrasable cavity techniques such as cryogenic cavities \cite{zhang2017ultrastable} and single-crystal mirror coatings \cite{kessler2012sub}, the use of multiple atomic references to pre-stabilize the local oscillator \cite{schioppo2017ultrastable}, or the use of active optical frequency references \cite{norcia2018frequency}.

For comparisons between multiple clocks making use of the same atomic transition, synchronized differential comparisons offer a way to bypass stringent requirements on the local oscillator. By probing two or more clocks with the same laser the effects of local oscillator noise can be cancelled out, offering coherent interrogation times well beyond the coherence time of the laser and eliminating the harmful Dick effect \cite{2022JunSr,zheng2022differential,clements2020lifetime}. This offers the prospect of high-stability differential comparisons between networks of optical clocks without the need for ultrastable cavities or frequency combs, which is promising for applications with demanding SWaP requirements such as space-based dark matter and gravitational wave detectors. 

Thanks to both continued improvements in local oscillator performance and the use of synchronous differential comparisons, clock comparisons are increasingly operating close to or at the quantum projection noise limit, also known as the standard quantum limit. This represents the fundamental limit on the achievable stability and precision of a clock comparison for atoms that are not in an entangled state, and scales as $\propto1/\sqrt{N_A}$, where $N_A$ is the atom number. Due to density broadening and shifts from atomic collisions, it is often difficult to significantly increase $N_A$ without harming clock accuracy or precision. Research into alternative clock geometries such as 3D optical lattice clocks \cite{campbell2017fermi} and tweezer clocks \cite{madjarov2019atomic,norcia2019seconds,young2020half} provides one promising route forward. Alternatively, the introduction of highly entangled quantum states such as spin-squeezed states or GHZ states offers the extremely appealing prospect of moving beyond the standard quantum limit scaling with $\propto1/\sqrt{N_A}$ to the ultimate Heisenberg limited scaling of $\propto1/N_A$, promising up to two orders of magnitude gain in stability for typical $N_A$ in optical lattice clocks. There has been considerable progress in this direction, including the recent generation of spin-squeezed states on optical clock transitions using a high-finesse optical cavity \cite{pedrozo2020entanglement}, and the demonstration of small scale optical GHZ states with Rydberg interactions in optical tweezers \cite{schine2021long}.

\textbf {Development of new atomic clocks based on highly charged ions (HCI)} --
HCIs have much higher sensitivities to the variation of  $\alpha$~\cite{kozlov_highly_2018,micke_coherent_2020} and
 are attractive candidates for the development of novel atomic clocks. Proposals based on HCI optical clocks and experimental progress towards HCI high-precision spectroscopy were recently reviewed in~\cite{kozlov_highly_2018}. Recent development of HCI cooling, trapping, and quantum logic techniques opened the door to rapid progress in the development of HCI clocks~\cite{micke_coherent_2020,King2021}. HCI clocks enable clock-comparisons with $\Delta K \approx 100$ \cite{2022QSNET}.
Since the demonstration of extreme-ultraviolet frequency combs (see, e.g., \cite{cingoz2012direct}), atomic clocks at wavelengths shorter than optical will became feasible presenting new opportunities for fundamental physics.

\textbf{ Development of a nuclear clock} --
Nuclear clock is based on a nuclear rather than an atomic transition~\cite{Peik2021}.
A long-lived nuclear transition that occurs between an excited state (isomer) of the $^{229}$Th isotope and the corresponding nuclear ground state, with wavelength near 150\,nm~\cite{2019nuclear} is within reach of modern lasers and enable development of a nuclear clock.
 The nuclear clock sensitivity to the variation of $\alpha$ is expected to exceed the sensitivity of present clocks by $\sim$4 orders of magnitude \cite{2020nuc}. It is highly sensitive to hadronic sector, with possible $K=10^4$ sensitivity to the variation of $m_q/\Lambda_\textrm{QCD}$ \cite{2020nuc}.
 In addition, nuclear clocks will be sensitive to a DM coupling to the hadronic sector of the SM.
 At the projected $10^{-19}$ fractional frequency precision level and strongly enhanced sensitivity, the nuclear clock can improve the ability to probe scalar dark matter by 5 - 6 orders of magnitude for a wide range of DM mass ranges in comparison with present limits~\cite{Kennedy:2020DM_atom-cavity}. The 229Th
nuclear optical clock represents a major challenge in view of the tremendous gap of nearly 17
orders of magnitude between the present uncertainty in the nuclear transition frequency (about $0.2~{\rm eV}$, corresponding to $\sim 48~{\rm THz}$) and the natural linewidth (in the mHz range). A detailed plan for the development of a nuclear clock is described in \cite{Peik2021}.
 
\textbf{Development of molecular clocks} --
The rotational and vibrational degrees of freedom in molecules involve motion of the nuclei themselves, while the chemical bond characteristics involve the electron mass. Thus molecules provide direct sensitivity to $m_p/m_e$ \cite{Patraeaba0453} and its variation. The best limit on this variation by use of molecules \cite{kobayashiNatureComm2019} remains three orders of magnitude less stringent than that using atoms \cite{2021YbclockAlpha}. Nonetheless, the atomic limit on $m_p/m_e$ relies on the Cs microwave clock and there are prospects for molecular clocks with optical-level sensitivities.  Optical transitions in trapped molecular ions \cite{2021Hanneke,carolloAtoms2018,kokishPRA2018} or lattice-confined neutral molecules \cite{zelevinskyPRL2008} are projected to have systematic and statistical limits comparable to those of the best optical atomic clocks. Accidental near-degeneracies between levels with differential sensitivity to $\alpha$- or $m_p/m_e$-variation are common in molecules and provide statistical enhancement -- allowing optical sensitivities in measurements that take place in the microwave. The use of such a near-degeneracy in KRb provides the current best limit in $m_p/m_e$ drifts by use of molecules \cite{kobayashiNatureComm2019}. Degeneracies have been identified in numerous species such as SrOH \cite{kozyryevPRA2021} and O$_2^+$ \cite{hannekePRA2016}. 

\subsection{Isotope shift precision measurements}
\label{Sec:3_IS}
Promising searches for new  particles are feasible through isotope-shift atomic spectroscopy, which is sensitive to a hypothetical fifth force between the neutrons of the nucleus and the electrons of the shell.
In so-called King plots, the mass-scaled frequency shifts of two optical transitions are plotted against each other for a series of isotopes~\cite{King:63}.
By considering the SM leading contribution to the IS (mass and field shifts), it was shown that there is a linear relation between two electronic transitions (with respect to different IS measurements).
New spin-independent interactions between the electron and the neutron will break this relation, and thus can be probed by looking for a deviation of the King plot from linearity~\cite{Delaunay:2016brc,Berengut:2017zuo}.
Therefore, the King plot analysis of precision isotope shift (IS) spectroscopy sets limits on spin-independent interactions that could be  mediated by a new particle which could be associated with dark matter.
These spin-dependent interactions can be a result of tree level exchanges of new scalar or vector bosons with a mass below $\mathcal{O}(50\,{\rm MeV})$, which appear in different extensions of the SM.
See Ref.~\cite{Frugiuele:2016rii} for interpretations in a context of several beyond the SM models.
Particularly interesting, in this regard, is a class of singlet scalar portal models, where the scalar mixes with the Higgs~\cite{OConnell:2006rsp}.
This class of models is motivated by the relaxion framework~\cite{Graham:2015cka} that effectively can be described as a finely tuned Higgs-portal model~\cite{Flacke:2016szy,Banerjee:2020kww}, and that may be potentially probed by the methods described here.

\textbf{R\&D: Fifth force searches with IS precision measurements} --
There are several  challenges in realizing full potential of this approach.
The most stringent bound on such fifth forces has been established based on Ca$^+$ isotope shift spectroscopy with up to 20\,Hz resolution~\cite{solaro_improved_2020}.
While Yb$^+$ measurements have revealed a $3\sigma$ nonlinearity~\cite{counts_observation_2020}, experiments in neutral Yb suggest that the nonlinearity may be caused by higher-order contributions within the SM~\cite{Figueroa2022Feb}. A possible improvement of the bounds can be achieved by choosing one transition in the King-plot comparison from a trapped-ion clock and the other from neutral atoms optical lattice clocks of the same species; e.g. Sr$^+$ and neutral Sr clocks.

Higher-order SM contributions break the linearity of the King plot as well~\cite{Berengut:2017zuo} and must either be calculated with high accuracy~\cite{yerokhin_nonlinear_2020}, or eliminated together with uncertainties on the isotope mass differences using more transitions and a generalized analysis~\cite{Mikami:2017ynz,Berengut:2020itu}.
Nuclear structure properties such as the quadrupole deformation and higher order nuclear moments can cause a non-linearity in the King plot. Thus, it is critical to understand the impact of nuclear effects to set constraints to BSM physics. As commonly only a few stable isotopes exist for a given element, adding points to a King plot requires measurements of unstable isotopes. Therefore, it is crucial to extend isotope shift measurements to unstable isotopes to be able to distinguish new physics from Standard Model effects \cite{Berengut:2020itu,counts_observation_2020}.

Isotope shift spectroscopy of highly charged ions, such as Ca$^{11+}$ to Ca$^{16+}$ isotopes is a promising avenue since the few-electron systems can be calculated with high accuracy and offer several narrow transitions for removing SM nonlinearities \cite{2021Ca}.


\subsection{Clock-based gravitational wave detection}
\label{sec:ClockGW}

Space-based optical lattice atomic clocks could potentially be used as gravitational wave detectors with unique capabilities \cite{2016clockGW}. The basic concept is similar to gravitational wave detection with space-based atomic interferometers, as discussed in Sec.~\ref{sec:AI}, with the primary difference being that in the case of optical lattice clocks the atoms remain strongly confined, necessitating the use of an on-board drag-free reference mass \cite{armano2018beyond}, but relaxing the requirements on atom temperature and satellite size, and enabling ground-based development and demonstration of required performance. High precision differential comparisons between optical lattice clocks in spatially separated satellites in inertial frames linked over an optical baseline eliminate sensitivity to local-oscillator frequency noise while remaining sensitive to Doppler shifts of the laser light induced by gravitational waves. This offers several potentially attractive features, including detectors consisting of only two spacecraft, relaxed requirements on the number of photons transmitted over the optical baseline relative to optical-interferometry-based detectors, and the capability to tune the detector response function using pulse sequences applied locally to the atoms on each spacecraft, without requiring any physical changes to the detector spatial configuration. The last feature in particular is of interest as it suggests the possibility of a tunable, narrowband GW detector that could lock onto and track specific GW signals from sources such as inspiraling intermediate and stellar black hole binaries as they leave the frequency band of space-borne optical detectors such as LISA \cite{amaro2017laser}, evolve through the decihertz band \cite{sedda2021missing}, and then enter the range of terrestrial optical-interferometry-based GW detectors such as LIGO \cite{abbott2016observation}.



\section{Fundamental physics with radioactive atoms and molecules}

Radioactive atoms and molecules offer extreme nuclear nuclear charge, mass, and deformations, and may be worked with efficiently with the advanced quantum control toolset of AMO. These rare systems offer an unprecedented amplification of both parity- and time-reversal violating properties. Hence, their study offers a wide range of opportunities for studies of nuclear structure, symmetry violations, and precision measurements. Here we discuss just a few motivating examples of the unique advantages of these species. 

\textbf{Symmetry violations}  The sensitivity of atoms and molecules to fundamental symmetry violations, such as P and CP, scales with the proton number of the nucleus as $\sim Z^{2-5}$ depending on the physical source.  Thus, the heaviest nuclei, which are necessarily radioactive, offer the largest sensitivity.  Furthermore, nuclei with octupole $(\beta_3)$ deformations and close proximity opposite parity nuclear states can enhance sources of hadronic CP violation by an additional factor of $10^{2-3}$ \cite{Haxton1983, Spevak1997}, making species such as Fr, Ra, Ac, Th, Pa very attractive for searches for CP-violation.  Molecules containing these nuclei can combine molecular and nuclear enhancements to offer extreme sensitivity, as much as $10^{5-6}$ per particle over state-of-the-art atomic experiments \cite{Graner2016}.  A recent major milestone towards this end include the creation and study of several radium-containing molecules which holds tremendous promise for symmetry violation searches  \cite{Gar2020,Yu2021, Fan2021}.  


\textbf{Clocks} There are unstable isotopes which provide a range of opportunities for optical clocks.  There has been a long-standing effort to realize a clock based on a nuclear isomeric transition in Th-229, which is unstable \cite{peik2003nuclear} as described in Section ~\ref{clocks}.  The radium ion is a candidate system for a transportable optical clock due to certain favorable properties, including photonic technology compatible transitions \cite{Holliman2022}.  It has been proposed to use a pair of clocks based on the highly charged ions Cf$^{+15}$ and Cf$^{+17}$ to search for the time variation of the fine structure constant \cite{Cf1}.

\textbf{Precision studies based on nuclear decays}  There are certain experiments where nuclear instability is essential.  For example, there is an experiment that aims to search for sterile neutrinos by reconstructing the kinematics of the radioactive decay of $^{131}$Cs atoms trapped in a MOT\cite{Hunter}.  There is also a recent proposal to test quantum mechanics by searching for an energy shift in an optical clock transition of a short-lived radioactive element \cite{Raizen:2022}. 

The study of radioactive isotopes pose several challenges, they are typically created at high-temperatures ($>$ 2000 $^\circ$C) and in the presence of large contaminants, with relative abundances that can be inferior to 1:10$^6$. Moreover, the lifetime of the isotopes of interest can be a few days or shorter. Therefore, their study demands particularly high efficiency and selectivity. Hence, the use of ion traps has proven to be \textbf{critical} for the study of radioactive species.  The beam-based methods used by the most sensitive ACME electron EDM experiment~\cite{andreev2018improved} requires macroscopic quantities of material, which is either impractical or impossible for many exotic nuclei.  Traps, on the other hand, can trade large count rates for longer coherence times, as demonstrated by the successful JILA HfF$^+$ eEDM \cite{Cairncross2017} and ANL $^{225}$Ra EDM experiments \cite{Parker2015}, the latter of which uses a radioisotope with a 15 day half-life and which cannot be obtained in macroscopic quantities.  Note that these two experiments use radically different trapping technologies, both of which are being actively developed but which require further development to fully realize the potential of exotic nuclei.  

\textbf{R \& D for ion traps} The JILA EDM experiment uses a molecular ion trap \cite{Cairncross2017}, an approach which can realize coherence times longer than one second, and therefore can make a competitive measurement with even tens of measured molecules per shot -- a combination that fits well with using short-lived and exotic species.  This technology which was recently used to synthesize, trap, and cool radioactive molecules including RaOCH$_3^+$ and RaOH$^+$, both of which are promising for nuclear CP-violation searches as they combine a deformed nucleus with polyatomic molecular advantages~\cite{Fan2021,Yu2021}.  Further development includes establishing measurement protocols for CP-violating observables, and utilizing methods to work with very short lived species.  To further increase efficiency it is desirable for state preparation and readout to utilize techniques such as quantum logic spectroscopy, which is nondestructive \cite{Chou2017}.  These experiments also require extensive molecular spectroscopy, as discussed below. 

\textbf{R \& D for neutral traps}  The ANL experiment uses an optical trap for neutral species, which offers similar coherence times as ion traps but with potentially much larger numbers.  However, traps for neutral species which are suitable for precision measurements tend to be very shallow, and therefore the species must be cooled to ultracold temperatures.  This presents a serious challenge for molecules due to their significant internal complexity, though there has been tremendous recent progress.  Since the first laser cooling of a diatomic molecule in 2010 \cite{Shuman2010}, the community has seen tremendous advances, including loading molecules into optical traps suitable for precision measurement~\cite{Anderegg2018,Anderegg2019}.  Furthermore, these techniques are now starting to be implemented in polyatomic molecules as well~\cite{Vilas2021}.  Polyatomic molecules offer unique advantages for CP-violation searches~\cite{Kozyryev2017PolyEDM,Yu2021,Hutzler2020Review}, but their complex structure makes them a challenge.  Further development includes improved and more efficient methods for production, slowing, and cooling molecules, has seen rapid recent development. Laser cooling schemes for radioactive molecules such as RaF have been investigated \cite{Gar2020}, but further developments are needed to produce ultracold molecules of these exotic species.  \\

All of these experiments also require extensive molecular spectroscopy, and in particular, dedicated facilities which can perform this spectroscopy with very short-lived species. 
These developments are well aligned with the progress of radioactive beam facilities around the world, such as the DOE facility for rare isotope beams (FRIB), where unique access to large amounts of actinide nuclei will be provided. Thus is critical to support the infrastructure needed to integrate table top experiments with radioactive beam facilities. This includes beam purification techniques, deceleration, cooling, and trapping of radioactive atoms and molecules.   

\bibliographystyle{apsrev4-1}
\bibliography{main.bib}

\end{document}